\documentclass[slac_one]{revtex4}
\usepackage{graphicx}
\usepackage{fancyhdr}
\begin{document}

\title{Exclusive measurements for SUSY events with
the ATLAS detector at the LHC} 

%

\author{A. Robichaud-V\'eronneau, on behalf of the ATLAS collaboration}
\affiliation{Universit\'{e} de Gen\`{e}ve, Geneva, Switzerland (email: andree.robichaud-veronneau@cern.ch)}

\begin{abstract}
We present recent work performed in ATLAS on techniques used to
reconstruct the decays of SUSY particles at the LHC. We concentrate on
strategies to be applied to the first fb$^{-1}$ of LHC data.
\end{abstract}

\maketitle

\thispagestyle{fancy}


\section{INTRODUCTION\label{intro}}
The Large Hadron Collider (LHC) has started operation very recently
and soon it will deliver $p-p$ collisions at a center-of-mass energy
of 14 TeV. The ATLAS detector will be used to search for evidence for
physics beyond the Standard Model (SM).

Among the many extensions to the SM that predict what this physics
might be, supersymmetry (SUSY) with R-parity conservation is a very
attractive one. It provides a candidate particle for dark
matter, the lightest neutralino, and predicts a light Higgs boson, in
agreement with electroweak precision measurements.

The following work is limited to the study of mSUGRA models. A list of
predefined points~\cite{CSCbook} in the parameter space is used. Since
there is no LHC data yet, events are generated with Isajet and Herwig,
and passed through a realistic simulation of the ATLAS detector. In
these models, pair production of SUSY particles is assumed, each
decaying in a cascade to the lightest supersymmetric particle (LSP),
which can only be detected by a missing transverse energy
signature.

\section{EDGE MEASUREMENTS}
Endpoint measurements are used when one particle is lost in the decay
or cannot be measured. In this case, the LSP is only detected by its
missing energy signature. To study this, the following
decay chain is used:
\begin{equation}\label{chain}
\widetilde{q}_L \to \widetilde{\chi}_2^0 q (\to \widetilde{l}^{\pm} l^{\mp} q ) \to  \widetilde{\chi}_1^0 l^+ l^- q
\end{equation}

This decay chain provides a large signal to background ratio due to
its final state. In the case of the ``Bulk'' point (SU3), the decay of
the neutralino goes through an extra step involving sleptons, since
$\widetilde{l}_R$ and $\widetilde{\tau}_1$ are lighter than
$\widetilde{\chi}_2^0$. For the ``Low Mass'' point (SU4), the
neutralino decays directly to a lepton pair and the LSP since sleptons
are heavier. The mSUGRA parameters for SU3 are m$_0$ = 100 GeV,
m$_{1/2}$ = 300 GeV, A$_0$ = -300 GeV, tan$\beta$ = 6, $\mu$ $>$ 0, and
for SU4, m$_0$ = 200 GeV, m$_{1/2}$ = 160 GeV, A$_0$ = -400 GeV,
tan$\beta$ = 10, $\mu$ $>$ 0. The NLO cross-section for SU3 is 27.68 pb
while for SU4, it is 402.19 pb.

\subsection{Dilepton edges}
By considering the lepton pair produced in eq.~\ref{chain}, it is
possible to obtain insights about the masses involved in the decay. In
the SU3 case, we have $m^{edge}_{ll} =
m_{\widetilde{\chi}_2^0}-m_{\widetilde{\chi}_1^0}$ and for SU4, the expression is more complex, as shown in eq. \ref{edge}.
The Events with two or three isolated leptons (electrons or muons) are
selected. Opposite sign (OS) lepton pairs are required in the
two-lepton events and all possible combinations of opposite sign
leptons are considered in the three-lepton events. Lepton pairs with
opposite-flavour (OF) are subtracted from the same-flavour (SF) pairs,
and cuts are performed on transverse missing energy
(E$_T^{\mathrm{miss}}$), transverse momenta of the four leading jets,
the ratio between E$_T^{\mathrm{miss}}$ and the effective mass and the
transverse sphericity.

\begin{equation}\label{edge}
m^{edge}_{ll} = m_{\widetilde{\chi}_2^0}
\sqrt{1-\Bigg(\frac{m_{\tilde{l}}}{m_{\widetilde{\chi}_2^0}}\Bigg)^2}
\sqrt{1-\Bigg(\frac{m_{\widetilde{\chi}_1^0}}{m_{\tilde{l}}}\Bigg)^2}
\end{equation}

\begin{figure}[!h]
\centering
\includegraphics[width=0.25\textwidth]{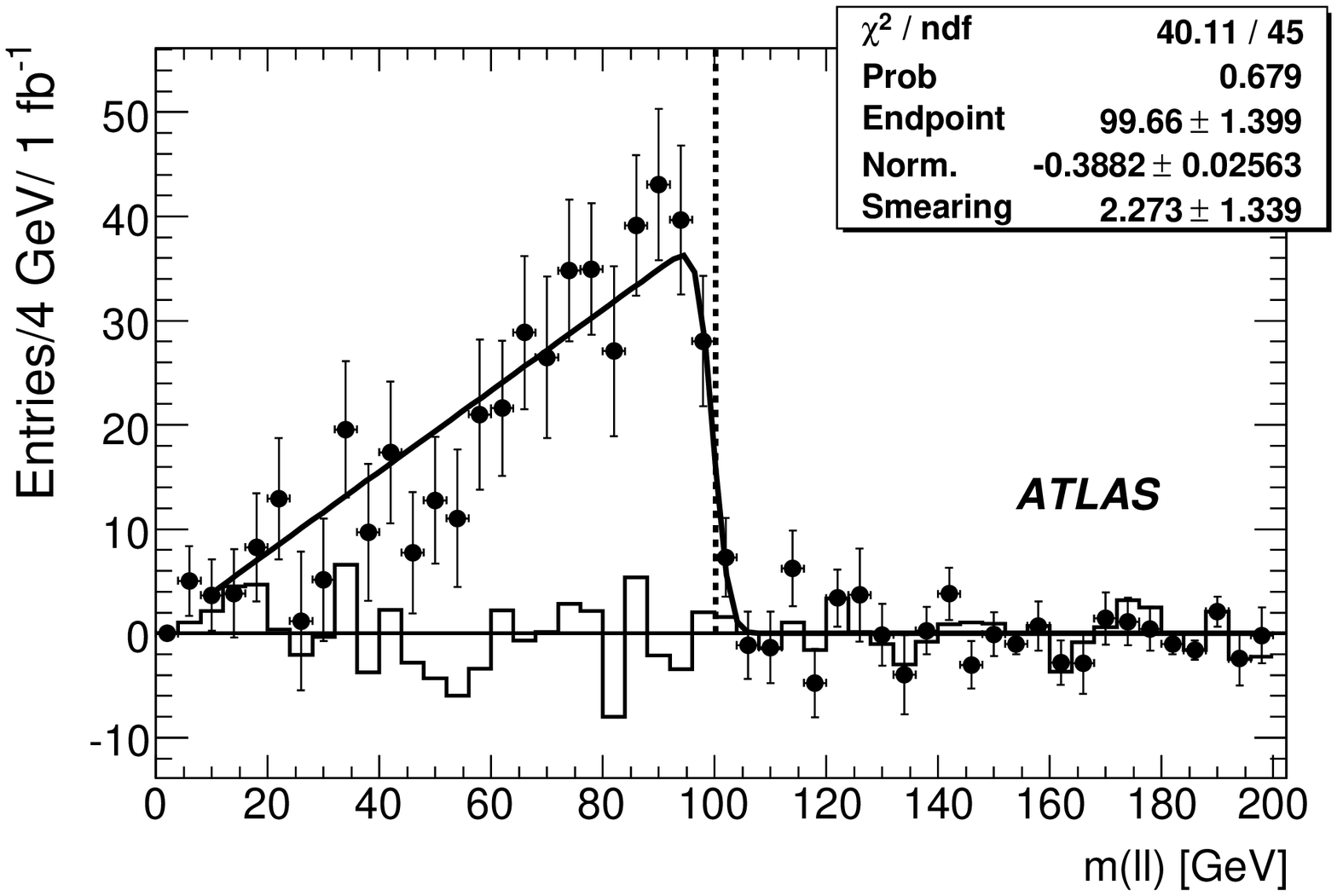}
\includegraphics[width=0.25\textwidth]{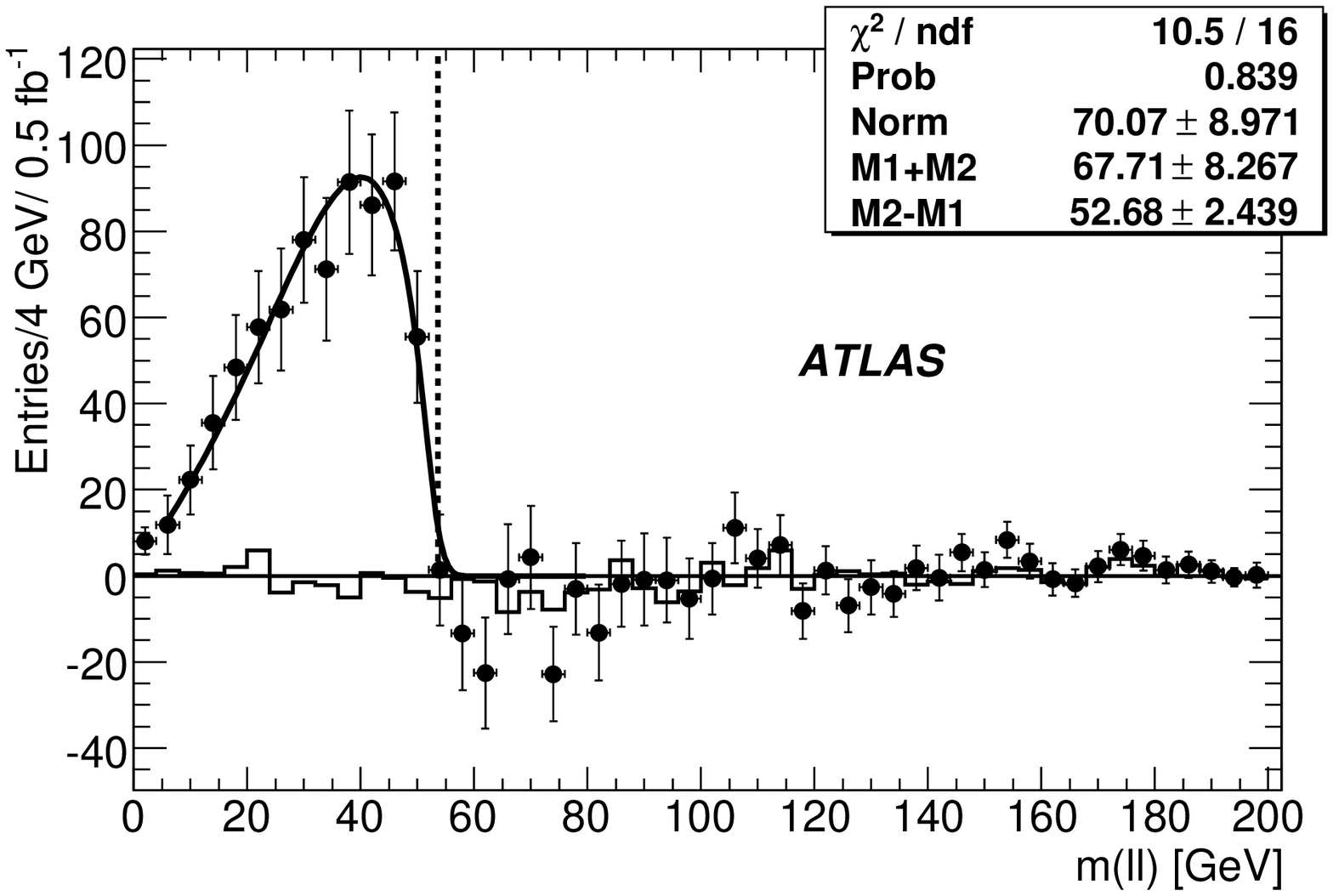}
\caption{Invariant mass distribution for SU3 (left) and SU4 (right). The points show the sum of the SM and the SUSY contributions, the line histogram is the SM contribution only. The result of the fit is superimposed and the dashed line show the expected position of the endpoint.}\label{edgefig}
\end{figure}

The invariant mass distribution is fitted with a triangular function
smeared with a Gaussian for the SU3 case for 1 fb$^{-1}$ as shown in
figure~\ref{edgefig}. The endpoint value obtained from the fit is
(99.7 $\pm$ 1.4 $\pm$ 0.3) GeV, where the quoted errors are
respectively the statistical error, the systematic error on the lepton
energy scale and the systematic error on the $\beta$
parameter~\cite{CSCbook}. The SU4 case requires a 3-body decay
theoretical distribution~\cite{3body} smeared for the experimental
resolution. The fit gives an endpoint of (52.7 $\pm$ 2.4 $\pm$ 0.2)
GeV for 0.5 fb$^{-1}$. The ``Coannihilation'' point (SU1) shows a
double edge in the same invariant mass distribution, due to both left-
and right-handed sleptons being lighter than
$\widetilde{\chi}_2^0$. The edges cannot be fitted with 1 fb$^{-1}$
although an excess is visible, while with 18 fb$^{-1}$, a fit can be
obtained with a lower edge at (55.8 $\pm$ 1.2 $\pm$ 0.2) GeV and a
upper edge at (99.3 $\pm$ 1.3 $\pm$ 0.3) GeV. All results are
consistent with the calculated values of 100.2 GeV (SU3), 53.6 GeV
(SU4) and 56.1 and 97.9 GeV (SU1).

\subsection{Jet + Lepton edges}
As can be seen in eq.~\ref{chain}, all masses can be reconstructed
using the jets in the final state to obtain endpoint
measurements. Three new quantities can be used: m$_{llq}$ (edge and
threshold), m$_{lq(high)}$ and m$_{lq(low)}$, which are the highest
and lowest value of m$_{lq}$ in an event using the same jet as
m$_{llq}$. Two straight lines, with a Gaussian smearing for a smooth
transition between them, are fitted to a small range of data points in
the m$_{llq}$ distribution, for both the edges and the thresholds. The
endpoints are explicitely fitted.
The results of the fits are shown in Table \ref{llq}.

\begin{table}[!h]
\centering
\begin{tabular}{|c||c|c||c|c|}
\hline
Endpoint & SU3 truth & SU3 measured & SU4 truth & SU4 measured \\
\hline
m$_{llq}^{max}$ & 501 & 517 $\pm$ 30 $\pm$ 10 $\pm$ 13 & 340 & 343 $\pm$ 12 $\pm$ 3 $\pm$ 9 \\
m$_{llq}^{min}$ & 249 & 265 $\pm$ 17 $\pm$ 15 $\pm$ 7 & 168 & 161 $\pm$ 36 $\pm$ 20 $\pm$ 4 \\
m$_{lq(low)}^{max}$ & 325 & 333 $\pm$ 6 $\pm$ 6 $\pm$ 8 & 240 & 201 $\pm$ 9 $\pm$ 3 $\pm$ 5 \\
m$_{lq(high)}^{max}$ & 418 & 445 $\pm$ 11 $\pm$ 11 $\pm$ 11 & 340 & 320 $\pm$ 8 $\pm$ 3 $\pm$ 8 \\
\hline
\end{tabular}
\caption{Endpoint positions from fits for SU3 (1 fb$^{-1}$) and SU4
(0.5 fb$^{-1}$), in GeV. Errors are respectively statistical,
systematic and jet energy scale uncertainty.}\label{llq}
\end{table}

\subsection{Tau signatures}
In the previous sections, leptons were considered to be only electrons
or muons. Tau leptons have to be treated separately. For the decay
$\widetilde{\chi}_2^0 \to \widetilde{\tau}^{\pm}_1 \tau^{\pm} \to
\widetilde{\chi}_1^0 \tau^+ \tau^- $, the branching ratio is 10 times
higher than for other leptons (for SU1 or SU3 scenarios). Also, since
the $\tau^{\pm}_1$ is involved in this decay and the neutralino masses
can be determined from other measurements, it is possible to determine
the $\tau^{\pm}_1$ mass. Since the decay of $\tau$ involves neutrinos
in the final state, it is not possible to get a sharp edge at the
maximum kinematic value.

Invariant mass distributions are plotted for SU1 and SU3 models, where
the same-sign (SS) distribution is subtracted from the opposite-sign
(OS) one. 
Special care is needed concerning the fit results due
to polarization effects on the $\tau$ invariant mass distribution,
which can considerably shift the position of the endpoint. 
 
\subsection{Right-handed squark pairs}
The decay chain presented in eq.~\ref{chain} holds for left-handed
squark decay. In the case of right-handed squarks, the decay goes
directly to the LSP and quark: $\widetilde{q}_R \to
\widetilde{\chi}_1^0 q$.  In this case, a new variable is introduced,
the ``stransverse mass'' m$_{T2}$~\cite{CSCbook}. Assuming the mass of
the LSP is known from previous measurements, m$_{T2}$ can be used to
determine the $\widetilde{q}_R$ mass. A linear fit is applied to a
range of data points around the edge of the m$_{T2}$ distribution to
determine the endpoint for SU3 and SU4 models. The results of the fit
are 591 $^{+13}_{-6}$ (\emph{sys}) $\pm$ 13 (\emph{stat}) GeV for SU3
and 407 $^{+10}_{-3}$ (\emph{sys}) $\pm$ 12 (\emph{stat}) GeV for SU4
These should be compared with the known values: 637 GeV for SU3 and
405 GeV for SU4.

\subsection{Light stop}

In the particular case of SU4, all SUSY masses are relatively light
and so is the $\widetilde{t}_1$, with a mass of 206 GeV. As it is
always decaying to the same channel, we can study the following:
$\widetilde{g} \to \widetilde{t}_1 t \to \widetilde{\chi}_1^{\pm} bt$
The upper endpoint of the \emph{tb} invariant mass depends on all
masses involved in the decay. Only the hadronic top decays are
included in the distribution. A fit is performed on the invariant mass
distribution (after W background is subtracted using the sideband
method) using a triangular function smeared with a Gaussian. It gives
a value for the endpoint of 297 $\pm$ 9 GeV (for a 5-parameter fit)
for 200 pb$^{-1}$, in agreement with the calculated value of $\sim$300
GeV.

\section{HIGGS IN SUSY EVENTS}

The Higgs boson can be produced in many ways at the LHC. Most
commonly, it is looked for in SM interactions (e.g. $g-g$ fusion), but
it can also occur in the decay of sparticles which were produced by
the initial interaction, like here for the neutralino in the SU9 model
(``Bulk'' point with enhanced Higgs production):
$\widetilde{\chi}_2^{0} \to \widetilde{\chi}_1^{0} h \to
\widetilde{\chi}_1^{0} b \bar{b}$. Requiring significant missing
transverse energy suppresses the QCD background, enabling the
observation of Higgs decay to b quarks.


\section{MASS AND PARAMETERS MEASUREMENT}

The different endpoint measurements obtained above for many mSUGRA
models can be used to determine the SUSY mass spectra and fits can be
performed to constrain the parameters of the given models. In some
cases (like the dilepton edges), an analytical formula is known to
describe the invariant mass shape and obtain the mass. In the other
cases, a $\chi^2$minimization procedure is used to obtain the
sparticle masses from several endpoint values. Parameters of the
mSUGRA models were obtained using 500 toy fits for both values of
sign($\mu$). For each fit, the observables are smeared using the full
correlation matrix. The results can be found in \cite{CSCbook}. For
the masses, there is agreement between theoretical and experimental
values, but the parabolic error for the minimization are still large.
As for the mSUGRA parameters, M$_0$ and M$_{1/2}$ can be determined
reliably, while in the case of tan $\beta$ and A$_0$, only the order
of magnitude can be obtained for the SU3 and SU4 models in a data
sample corresponding to 1 fb$^{-1}$.


\begin{acknowledgments}
This work is a collaborative effort of the ATLAS SUSY Working Group and many thanks go to them.
\end{acknowledgments}


\begin{thebibliography}{9}   


\bibitem{CSCbook}
ATLAS Collaboration,``Expected Performance of the ATLAS Experiment, Detector, Trigger and Physics'', CERN-OPEN-2008-020, Geneva, 2008, to appear.

\bibitem{3body}
U. De Sanctis, T. Lari, S. Montesano, C. Troncon, Eur. Phys. J. C {\bf52} (2007) 743. 


\end{thebibliography}
\end{document}